\documentclass[twocolumn,floatfix,preprintnumbers,nofootinbib,superscriptaddress]{revtex4}

\usepackage{ulem}
\usepackage{bm}
\usepackage{times}
\usepackage{amssymb,amsbsy,amsmath,amsfonts}
\usepackage{graphicx}
\usepackage{float}
\usepackage{color}
\usepackage{morefloats}
\usepackage{rotating}
\usepackage{srcltx}
\usepackage{slashed}
\usepackage{subfigure}
\usepackage{multirow}
\usepackage{verbatim}
\usepackage{hyperref}
\usepackage{tabularx}

\usepackage[outdir=./]{epstopdf}

\begin{document}

\title{Further understanding the nature of $a_0(1710)$ in the $D^+_s \to \pi^0 K^+ K^0_S$ decay}

\author{Xin Zhu}
\affiliation{School of Physics and Microelectronics, Zhengzhou
University, Zhengzhou, Henan 450001, China} \affiliation{Institute
of Modern Physics, Chinese Academy of Sciences, Lanzhou 730000,
China}
\author{Hao-Nan Wang}
\affiliation{Guangdong Provincial Key Laboratory of Nuclear Science, Institute of Quantum Matter,
South China Normal University, Guangzhou 510006, China}
\affiliation{Institute of Modern Physics, Chinese Academy of Sciences, Lanzhou 730000, China}
\author{De-Min Li}~\email{lidm@zzu.edu.cn}
\affiliation{School of Physics and Microelectronics, Zhengzhou
University, Zhengzhou, Henan 450001, China}
\author{En Wang}~\email{wangen@zzu.edu.cn}
\affiliation{School of Physics and Microelectronics, Zhengzhou
University, Zhengzhou, Henan 450001, China}
\author{Li-Sheng Geng}~\email{lisheng.geng@buaa.edu.cn}
\affiliation{School of Physics, Beihang University, Beijing 102206,
China} 
\affiliation{Peng Huanwu Collaborative Center for Research and Education, Beihang University, Beijing 100191, China}
\affiliation{Beijing Key Laboratory of Advanced Nuclear
Materials and Physics, Beihang University, Beijing 102206, China}
\author{Ju-Jun Xie}~\email{xiejujun@impcas.ac.cn}
\affiliation{Institute of Modern Physics, Chinese Academy of
Sciences, Lanzhou 730000, China} \affiliation{School of Nuclear
Sciences and Technology, University of Chinese Academy of Sciences,
Beijing 101408, China}

\date{\today}

\begin{abstract}

Based on our previous work about the role of $a_0(1710)$   in the $D_s^+\to\pi^+K_S^0K_S^0$ decay [Phy. Rev. D 105, 116010 (2022)], we perform a further theoretical study of $a_0(1710)^+$  in the process $D^+_s \to \pi^0 a_0(1710)^+ \to \pi^0 K^+ K^0_S$. In addition to $a_0(1710)$, the contributions of $K^*$ and $a_0(980)$ are also taken into account. Firstly, we consider the contributions from the tree diagrams of $K^{*+} \to K^+\pi^0$ and $\bar{K}^{*0} \to \pi^0 \bar{K}^0$. Secondly, we describe the final state interaction of $K\bar{K}$ in the chiral unitary approach to study the contribution of $a_0(980)$, while the $a_0(1710)$  state is dynamically generated from the $K^*\bar{K}^*$  interaction, and then decays into $K^+\bar{K}^0$. Since the final $K^+ K_S^0$ state is in pure isospin $I=1$, the $D_s^+\to\pi^0K^+K_s^0$ decay is an ideal process to study the $a_0(1710)^+$ and $a_0(980)^+$ resonances. Based on our theoretical calculations, it is found that the recent experimental measurements on the $K^+K^0_S$, $\pi^0K^+$, and $\pi^0 K_S^0$ invariant mass distributions can be well reproduced, which supports the molecular $K^*\bar{K}^*$ nature of the scalar $a_0(1710)$ resonance.  

\end{abstract}

\maketitle

\section{Introduction} \label{sec:Introduction}

The $a_0(1710)$ resonance, an isospin partner of the scalar meson $f_0(1710)$, may have  properties and structure similar to $f_0(1710)$~\cite{Zhu:2022wzk,Dai:2021owu,Geng:2008gx,Geng:2009gb,Du:2018gyn,Wang:2021jub,Wang:2022pin}. The $f_0(1710)$ resonance, with quantum numbers $J^{PC} = 0^{++}$, has been investigated in many previous theoretical works~\cite{Workman:2022ynf,Nagahiro:2008bn,Branz:2009cv,Geng:2010kma,Wang:2011tm,MartinezTorres:2012du,Xie:2014gla,Dai:2015cwa,Dai:2018thd,Molina:2019wjj,Garcia-Recio:2013uva}, where its structure has been studied from various perspectives. For example, in Ref.~\cite{Close:2005vf} it was shown that the $f_0(1710)$ resonance has a large component of $s\bar{s}$ in its wave function. In Refs.~\cite{Gui:2012gx,Janowski:2014ppa,Fariborz:2015dou} it was considered as a scalar glueball. On the other hand, the $f_0(1710)$ can be viewed as a  molecular state dynamically generated in the vector-vector coupled channel interactions~\cite{Nagahiro:2008bn,Branz:2009cv,Geng:2010kma,Wang:2011tm,MartinezTorres:2012du,Xie:2014gla,Dai:2015cwa,Dai:2018thd,Molina:2019wjj}. Meanwhile, the $a_0(1710)$ with negative $G$-parity is also dynamically generated in the isospin $I=1$ sector~\cite{Geng:2008gx,Geng:2009gb,Du:2018gyn}, which couples mostly to the $K^*\bar{K}^*$ channel, as well as to the $\rho \omega$ and $\rho \phi$ channels. In Ref.~\cite{Wang:2022pin}, after extending the coupled channel vector meson-vector meson interactions to include the pseudoscalar meson-pseudoscalar meson interactions, a pole near the $K^*\bar{K}^*$ threshold identified as the $a_0(1710)$ state is also found.

In the molecular picture, the $f_0(1710)$ and $a_0(1710)$ states couple mostly to the $K^*\bar{K}^*$ channel and their dominant decay channel is $K\bar{K}$~\cite{Geng:2008gx,Wang:2022pin}. This is very similar to the case of scalar mesons $f_0(980)$ and $a_0(980)$~\cite{Oller:1997ti,Oller:1997ng,Oller:1998hw}, which are dynamically generated  in the coupled $K\bar{K}$, $\pi\eta$, and $\pi\pi$ channels. The dominant decay channel of $f_0(980)$ is $\pi\pi$, while it is the $\pi\eta$ channel for $a_0(980)$.

On the experimental side, the BESIII experiment reported the first evidence for the interference between $f_0(1710)$ and $a_0(1710)$ in the amplitude analyses of the $D_s^+\rightarrow\pi^+K_S^0K_S^0$ and $D_s^+\rightarrow\pi^+K^+K^-$ decays~\cite{BESIII:2020ctr,BESIII:2021anf}. It is found that there is a clear enhancement in the $K^0_SK^0_S$ invariant mass distributions around $1.7$ GeV, which indicates the contribution from hte $a_0(1710)$ state. However, it is not seen in the $K^+K^-$ invariant spectrum. In fact, the $a_0(1710)^+$ state was previously reported by the $BABAR$ experiment~\cite{BaBar:2021fkz} in the process of $\eta_c 
\to a_0(1710)^+ \pi^-$ with the decay $a_0(1710)^+ \to \pi^+ \eta $. Its measured mass and width are $M_{a_0(1710)} = 1704 \pm 5 \pm 2 ~{\rm MeV}$ and $\Gamma_{a_0(1710)} = 110 \pm 15 \pm 11 ~{\rm MeV}$, respectively~\cite{BaBar:2021fkz}.

On the theoretical side, within the chiral unitary approach, the productions of the $f_0(1710)$ and $a_0(1710)$ states in $D_s^+\rightarrow\pi^+K_S^0K_S^0$ and $D_s^+\rightarrow\pi^+K^+K^-$ reactions can be well explained~\cite{Dai:2021owu}. While in our previous work~\cite{Zhu:2022wzk}, for the decay $D_s^+\rightarrow\pi^+K_S^0K_S^0$, we found that the contribution from the $K^*$ meson is crucial to the $a_0(1710)$ peak region. Furthermore, within the proposed mechanisms in Refs.~\cite{Zhu:2022wzk,Dai:2021owu} it is expected that the charged $a_0(1710)^+$ resonance will show up in the $K^+K^0_S$ invariant mass distribution of the $D^+_s \to \pi^0 K^+K^0_S$ decay. Recently, the BESIII Collaboration has performed an amplitude analysis
of the process $D_s^+ \to \pi^0 K^+K^0_S$~\cite{BESIII:2022wkv}, where the $a^+_0(1710)$ state was indeed observed in the $K^+K^0_S$ invariant mass distribution, and presented  the fitted Breit-Wigner mass and width as,
\begin{eqnarray}
 M_{a_0(1710)} &=& 1817 \pm 8 \pm 20 ~{\rm MeV},  \nonumber \\
 \Gamma_{a_0(1710)} &=& 97 \pm 22 \pm 15 ~{\rm MeV}.
\end{eqnarray}
Here, the fitted mass is close to the boundary region of the $K^+\bar{K}^0$ invariant mass spectrum.\footnote{If we take $M_{D^+_s} = 1968.34$ MeV, and $m_{\pi^0} = 134.98$ MeV, we get the maximum value of the $K^+\bar{K}^0$ invariant mass  $(M_{K^+\bar{K}^0})_{\rm max} = 1833.36$~MeV.} On the other hand, the contribution from the $K^*$ meson is also significant to the  $a_0(1710)$ peak region. In the low energy region of the $K^+\bar{K}^0$ line shape, there is a clear enhancement due to the scalar meson $a_0(980)$.

Based on the BESIII experimental measurements ~\cite{BESIII:2022wkv}, Ref.~\cite{Guo:2022xqu} has systematically studied these isovector scalar mesons, and suggested that the $a_0(1710)$ [denoted as $a_0(1817)$], $a_0(1450)$, and $a_0(980)$ states are in the same isovector scalar meson family since they form a Regge trajectory. In addition, the $a_0(1710)$ state could also be regarded as a good isovector partner of the $X(1812)$~\cite{Guo:2022xqu}. 
In a word, the nature of $a_0(1710)$ is still unclear, and the further analysis of the experimental data is necessary.

In this work, following these previous theoretical works in Refs.~\cite{Zhu:2022wzk,Xie:2014tma,Molina:2019udw,Wang:2021naf,Duan:2020vye,Ling:2021qzl} we investigate the roles of the scalar mesons $a_0(980)^+$ and $a_0(1710)^+$ in the $D_s^+ \to \pi^0 K^+K^0_S$ decay, where their contributions are included by taking into account the $K\bar{K}$ and  $K^*\bar{K}^*$ final-state interactions within the chiral unitary approach. In addition, we also consider the contributions of the intermediate states $K^{\ast+}$ and $\bar{K}^{\ast 0}$ in the processes $D^+_s \to \bar{K}^0 K^{*+} \to \pi^0 K^+K^0_S$ and $D^+_s \to K^+ \bar{K}^{*0} \to \pi^0 K^+K^0_S$.

This article is organized as follows. In Sec.~II, we present the theoretical formalism for the  $D_s^+ \to \pi^0 K^+ K^0_S$ decay, and in Sec.~III, we show our theoretical numerical results and discussions, followed by a short summary in the last section.

\section{Formalism and Ingredients} 

In this section, we present the formalism and ingredients for the decay $D^+_s \to \pi^0 K^+ K^0_S$, where we study the roles of the $K^{*+}$ ($\bar{K}^{*0}$), $a_0(980)$, and $a_0(1710)$ states. The contribution of the $a_0(980)$ state is encoded in the $S$-wave $K\bar{K}$ final-state interactions, while the one of the $a_0(1710)^+$ state is in the $S$-wave $K^{\ast}\bar{K}^{\ast}$ final-state interactions.

\subsection{The mechanism of $D^+_s \to \bar{K}^0 K^{*+} (K^+\bar{K}^{*0}) \to \pi^0 K^+K^0_S$ reaction}

\begin{figure}[htbp]
	\centering
	\includegraphics[scale=0.55]{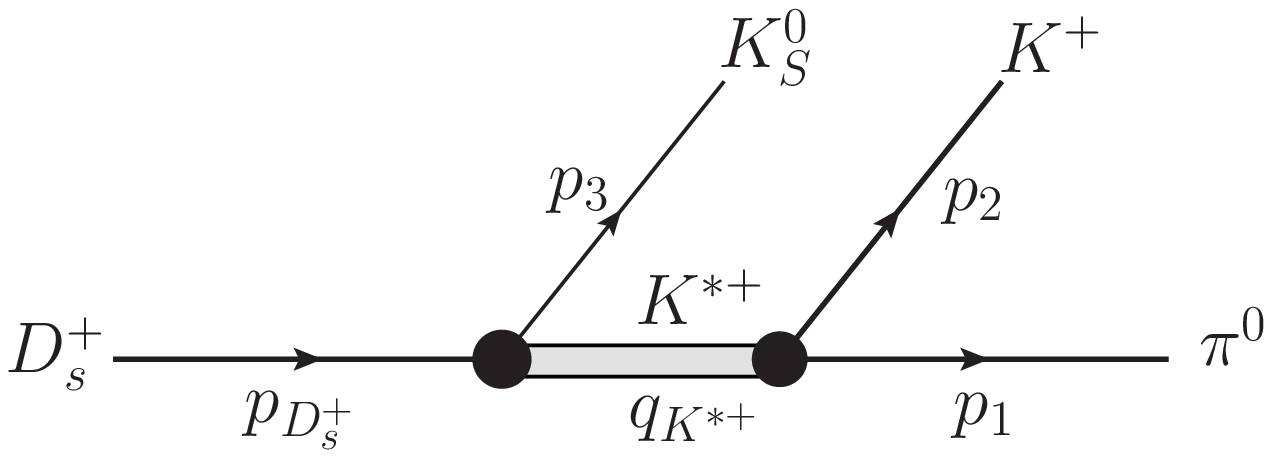}
	\caption{The decay of $D_s^+\rightarrow\pi^0K^+K_S^0$ via the intermediate vector $K^{\ast+}$. 
 }	\label{fig:mb}
\end{figure}

Firstly, we study the decay of $D_s^+ \to \pi^0K^+K_S^0$ via the intermediate resonance $K^{*+}$, with $K^{*+} \to \pi^0 K^+$ in $P$-wave. The vertex $D^+_s\bar{K}^0K^{*+}$ is also in $P$-wave to keep the angular momentum conserved, and the effective interaction is taken as used in Refs.~\cite{Zhu:2022wzk,Ling:2021qzl,Hsiao:2019ait}. The hadron level diagram for the process $D^+_s \to \bar{K}^0 K^{*+} \to \pi^0 K^+K^0_S$ is depicted in Fig.~\ref{fig:mb}, and the decay amplitude can be written as
\begin{eqnarray}
    \mathcal{M}_{K^{*+}} &=&\frac{g_{D_s^+\bar{K}^0K^{*+}}g_{K^{\ast+}K^+\pi^0}}{q^2_{K^{*+}}-m_{K^{\ast+}}^2+im_{K^{\ast+}}\Gamma_{K^{\ast+}}} \times \nonumber \\
    && \Bigg[\left(m_{K^+}^2-m_{\pi^0}^2\right)\left(1-\frac{q^2_{K^{*+}}}{m_{K^{\ast+}}^2}\right) + \nonumber \\
    && 2p_1\cdot p_3 \frac{m_{\pi^0}^2-m_{K^+}^2-m_{K^{\ast+}}^2}{m_{K^{\ast+}}^2} + \nonumber \\ 
&& 2p_2 \cdot p_3 \frac{m_{\pi^0}^2-m_{K^+}^2+m_{K^{\ast+}}^2}{m_{K^{\ast+}}^2} \Bigg], 
\end{eqnarray}
where $q_{K^{*+}}^2 = (p_1 + p_2)^2 = M_{{\pi^0}K^+}^2$ is the four momenta square of the virtual  $K^{\ast+}$ meson, and $M_{{\pi^0}K^+}$ is the invariant mass of the $\pi^0 K^+$ system.

\begin{figure}[htbp]
	\centering
	\includegraphics[scale=0.55]{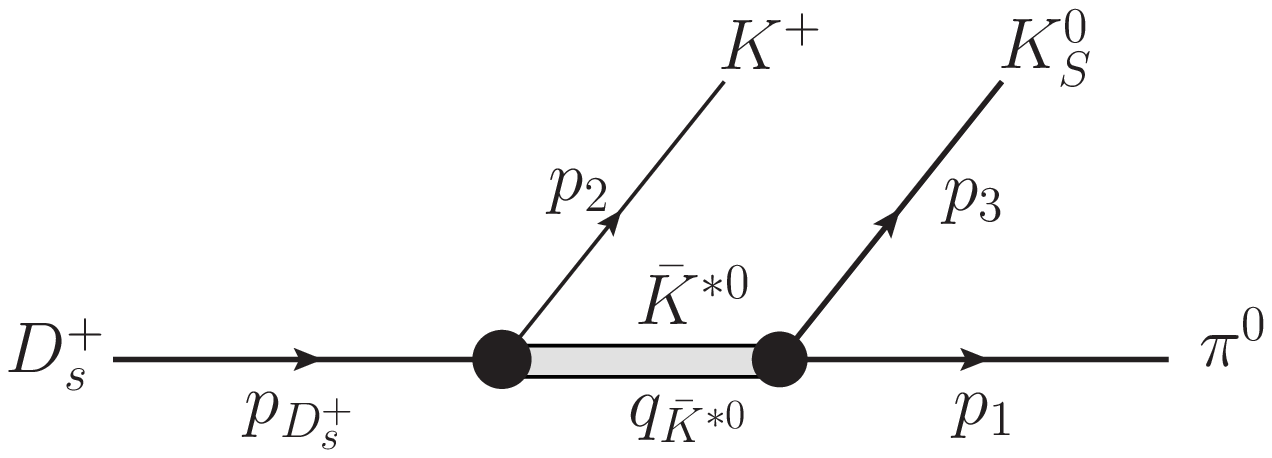}
	\caption{The decay of $D_s^+\rightarrow\pi^0K^+K_S^0$ via the intermediate vector $\bar{K}^{\ast0}$. 
 }	\label{fig:mc}
\end{figure}

On the other hand, as shown in Fig.~\ref{fig:mc}, the decay of $D_s^+ \to \pi^0K^+K_S^0$ can also occur via the intermediate meson $\bar{K}^{*0}$. The corresponding decay amplitude for the process  of $D^+_s \to \pi^0 K^+K^0_S$  of Fig.~\ref{fig:mc} can be easily obtained just by applying the substitution to ${\cal M}_{K^{*+}}$ with $p_2 \leftrightarrow p_3$, $K^{*+} \to \bar{K}^{*0}$, and $q_{K^{*+}} \to q_{\bar{K}^{*0}}$.

The coupling constants $g_{D_s^+\bar{K}^0K^{*+}}$, $g_{K^{*+}K^{+} \pi^0}$, $g_{D_s^+\bar{K}^{\ast0}K^{+}}$, and $g_{\bar{K}^{*0}\bar{K}^0 \pi^0}$ are determined from the experimental partial decay widths of $D^+_s\to \bar{K}^0 K^{*+}$, $K^{*+} \to K^+\pi^0$, $D^+_s\to K^+ \bar{K}^{*0}$, and $\bar{K}^{*0} \to \bar{K}^0\pi^0$, respectively. The  results are listed in Table~\ref{tab:particleparameters}. The coupling constants are taken as real and positive as done in Refs.~\cite{Zhu:2022wzk,Ling:2021qzl}. In our study, the partial decay width of $K^* \to K\pi$ is taken from the review of particle physics (RPP)~\cite{Workman:2022ynf}. While for the $D^+_s \to K^{*+} \bar{K}^{0}$ dcay, we take $\mathrm{Br}(D^+_s\to \bar{K}^0 K^{*+}) = (5.4\pm1.2)\%$  from Ref.~\cite{Workman:2022ynf}, then one can easily obtain $\mathrm{Br}(D^+_s\to K^+ \bar{K}^{*0}) = (12.7\pm2.8)\%$ with the recent BESIII measurement of $\frac{\mathrm{Br}(D_s^+\to \bar{K}^{\ast0} K^+)}{\mathrm{Br}(D_s^+\to \bar{K}^0K^{\ast+})}=2.35^{+0.42}_{-0.23_{\text{stat.}}} \pm 0.10_{\text{syst.}}$ \cite{BESIII:2022wkv}.

\begin{table*}[htbp]
\renewcommand\arraystretch{1.2}
\caption{Four two-body decays and the corresponding coupling constants.}	\label{tab:particleparameters}
	\begin{tabular}{cccc}
		\hline\hline  
		~~~~Decay process~~~~ & ~~~~Partial decay width (MeV)~~~~ & ~~~~Coupling constant~~~~ & ~~~~Value~~~~ \\ \hline
		$D^+_s\to \bar{K}^0 K^{*+}$  &  $(0.71 \pm 0.16) \times10^{-10}$  & $g_{D_s^+\bar{K}^0K^{*+}}$ & $ (1.05 \pm 0.12) \times 10^{-6}$             \\
    	$K^{*+} \to K^+\pi^0$  & $16.9 \pm 0.3$  & $g_{K^{*+}K^{+} \pi^0}$ & $3.23\pm0.03$           \\
		$D^+_s\to K^+ \bar{K}^{*0}$  &  $(1.66\pm0.37)\times10^{-10}$  & $g_{D_s^+\bar{K}^{\ast0}K^{+}} $ & $(1.62 \pm 0.18)\times 10^{-6}$           \\
		$\bar{K}^{*0} \to \bar{K}^0\pi^0$  &  $15.8 \pm 0.2$   & $g_{\bar{K}^{\ast0}\bar{K}^0\pi^0}$ & $3.13\pm0.02$  \\
		\hline\hline
	\end{tabular}
\end{table*}

\subsection{The mechanism of $D^+_s \to \pi^0 a_0(980)^+ \to \pi^0 K^+K^0_S$ reaction}

The Cabibbo-favored process $D^+_s \to \pi^0 K^+ \bar{K}^0$ can happen via the weak decay of the $c$ quark into a $W^+$ boson and an $s$ quark, followed by the $W^+$ boson decaying into a $u$ quark and a $\bar{d}$ quark. In order to generate  the states $\pi^0 K^{(*)+}\bar{K}^{(*)0}$, the $u\bar{s}$ ($s\bar{d}$)  hadronize into $K^{(*)+}$ [$\bar{K}^{(*)0}$], and the $s\bar{d}$ ($u\bar{s}$), together with the $q\bar{q}(=u\bar{u}+d\bar{d}+s\bar{s})$ pair created from the vacuum, hadronize into $\pi^0\bar{K}^{(*)0}$ [$\pi^0K^{(*)+}$]. Then the scalar meson $a_0(980)$ $[a_0(1710)]$ could be dynamically generated from the $S$-wave $K^{(*)+}\bar{K}^{(*)0}$ interaction.

For the production of the $a_0(980)$ resonance, we need to produce the $\pi^0K^+\bar{K}^0$ in the first step, and then the $K^{+} \bar{K}^{0}$ final-state interaction generates the $a_0(980)$ state. Following Ref.~\cite{Molina:2019udw}, for the hadronization of $s\bar{d}$ and $u\bar{s}$ into a pair of pseudoscalar mesons, we can write: 

\begin{eqnarray}
\sum_{i=u,d,s}{s\bar{q}_iq_i\bar{d}} &=& M_{3i}M_{i2} = (M^2)_{32},  \label{eq:sdbarhadronization}\\
\sum_{i=u,d,s}{u\bar{q}_iq_i\bar{s}} &=& M_{1i}M_{i3} = (M^2)_{13},  \label{eq:usbarhadronization}
\end{eqnarray}
where $M$ is the $SU(3)$ quark pair $q_i\bar{q}_j$ matrix, and it is defined as~\cite{Geng:2010kma,MartinezTorres:2009uk}
\begin{eqnarray}
	M=\left(\begin{matrix} u\bar{u} & u\bar{d} & u\bar{s}  \\
		d\bar{u}  &   d\bar{d}  &  d\bar{s} \\
		s\bar{u}  &  s\bar{d}   &    s\bar{s}
	\end{matrix}
	\right).
\end{eqnarray}
Accordingly, at the hadron level, if we consider the $\eta$ and $\eta'$ mixing as in Ref.~\cite{Bramon:1992kr}, we can write the matrix $M$ in terms of the pseudoscalar ($P$) mesons,
\begin{eqnarray}
P =\left(\begin{matrix} \frac{\eta}{\sqrt{3}}+ \frac{{\pi}^0}{\sqrt{2}}+ \frac{{\eta}'}{\sqrt{6}} & \pi^+ & K^+  \\
    \pi^-  &   \frac{\eta}{\sqrt{3}}- \frac{{\pi}^0}{\sqrt{2}}+ \frac{{\eta}'}{\sqrt{6}}  &  K^0 \\
    K^-  &  \bar{K}^{0}   &    -\frac{\eta}{\sqrt{3}}+ \frac{{\sqrt{6}\eta}'}{3}
\end{matrix}
\right).
\end{eqnarray}

Then, the hadronization processes at the quark level in Eqs.~\eqref{eq:sdbarhadronization} and \eqref{eq:usbarhadronization} can be expressed at the hadron level as, 
 \begin{eqnarray}
(M^2)_{32} \to (P \cdot P)_{32} &=& \pi^+K^{-}-\frac{1}{\sqrt{2}}\pi^0\bar{K}^{0}, \\
(M^2)_{13} \to (P \cdot P)_{13} &=& \pi^+K^{0}+\frac{1}{\sqrt{2}}\pi^0K^{+}. 
\end{eqnarray}

\begin{figure}[htbp]
\subfigure[]
{\includegraphics[scale=0.65]{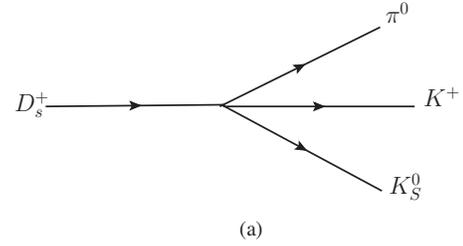}}
\subfigure[]
{\includegraphics[scale=0.52]{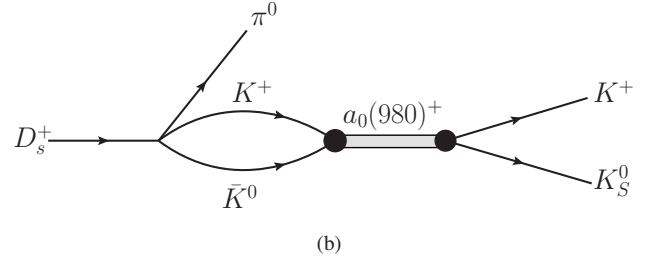}}
\caption{The mechanisms of the $D_s^+\rightarrow\pi^0K^+K_S^0$ decay. (a): tree diagram and (b): the final-state interaction of $K^+\bar{K}^0$ to produce the $a_0(980)^+$ state.} \label{Fig:md}
\end{figure}

Together with the other final state $K^+$ (hadronizatized directly from $u\bar{s}$) or $\bar{K}^0$ (hadronizatized directly from $s\bar{d}$), we obtain the $D^+_s \to \pi^0K^+\bar{K}^0$ reaction at tree level [shown in Fig.~\ref{Fig:md} (a)] with the above mechanism as: 
\begin{eqnarray}
M^{\rm tree}_{\pi^0 K^{+}\bar{K}^{0}} 
  &=& \frac{V_{\pi^0K^+\bar{K}^0}}{\sqrt{2}}, \label{eq:usbarhadron}
\end{eqnarray}
where we take $V_{\pi^0K^+\bar{K}^0}$ as a constant, as in most previous theoretical works~\cite{Liang:2014tia,Oset:2016lyh,Xie:2016evi,Wang:2022nac,Dai:2015bcc,Xie:2018rqv}.

Next, the final-state interaction of $K^+\bar{K}^0$ will produce the $a_0(980)$ state. The corresponding diagram is shown in Fig.~\ref{Fig:md} (b). Thus, the decay amplitude for the  $D_s^+ \to \pi^0K^+K_S^0 $ reaction shown in Fig~\ref{Fig:md} can be easily obtained as follows,~\footnote{It is worth mentioning that we have neglected the the $\pi K$ final-state interactions, where the scalar meson $\kappa(700)$ appears. This is because that its contribution is very small compared with the dominant contribution from the $K^*$ meson.}
 \begin{eqnarray}
\mathcal{M}_{a_0(980)}  &=& \frac{V_{\pi^0K^+\bar{K}^0}}{\sqrt{2}} \left[ 1 + G_{K\bar{K}}(M_{K^+\bar{K}^0}) \times \right. \nonumber \\
&& \left. t_{K^+\bar{K}^0 \to K^+\bar{K}^0}(M_{K^+\bar{K}^0}) \right],
 \label{eq:md}
\end{eqnarray}
with $M_{K^+\bar{K}^0}$ the invariant mass of the $K^+\bar{K}^0$ system.

In  Eq.~\eqref{eq:md}, $G_{K\bar{K}}$ is the $K\bar{K}$ loop function, which is given by,
\begin{eqnarray}
G_i = i \int \frac{d^4q}{(2\pi)^4}\frac{1}{(P-q)^2-m_1^2+i\epsilon}\frac{1}{q^2-m_2^2+i\epsilon},\label{eq:G}
\end{eqnarray}
where $m_1$ and $m_2$ are the masses of the two mesons in the loop of the $i$th channel, and $P$ and $q$ are the four-momenta of the two-meson system and the second meson, respectively. The loop function of Eq.~\eqref{eq:G} is logarithmically divergent. There are two methods to solve this singular integral, either using the three-momentum cut-off method, or the dimensional regularization method. In our work, we adopt the cut-off method, and  perform the integral for $q$ in Eq.~\eqref{eq:G} with a cut-off $|\vec{q}_{\rm max}|=903$ MeV to regularize the loops (see Ref.~\cite{Liang:2013yta} for more details).

On the other hand, the transition amplitude $t_{K^+\bar{K}^0 \to K^+\bar{K}^0}$, which is dependent on the invariant mass $M_{K^+\bar{K}^0}$, can be obtained in the chiral unitary approach by solving the Bethe-Salepter equation~\cite{Oller:1997ng},
\begin{eqnarray}
T&=&[1-VG]^{-1}V,
\end{eqnarray}
where $V$ is a 2$\times$ 2 matrix with the transition potential between the isospin channels $K\bar{K}$ and $\pi\eta$. With the isospin multiplets $K=(K^+,{K}^0),\bar {K}=(\bar{K}^0,-K^-)$, and $\pi=(-\pi^+,\pi^0,\pi^-)$, the $2\times2$ matrix $V$ can be easily obtained as follows,
\begin{eqnarray}
V_{K\bar{K}\to K\bar{K} } &=& -\frac{1}{4f^2}s,\nonumber \\
V_{K\bar{K}\to\pi \eta } &=& \frac{\sqrt{6}}{12f^2}(3s-\frac{8}{3}m_K^2-\frac{1}{3}m_\pi^2-m_\eta^2),\nonumber \\
V_{\pi \eta \to K\bar{K}} &=& V_{K\bar{K}\to\pi \eta }, \nonumber \\
V_{\pi \eta\to\pi \eta} &=& -\frac{1}{3f^2}m_\pi^2,  
\end{eqnarray}
where $f = 93$ MeV is the pion decay constant, and $s$ is the invariant mass squared of the pseduoscalar-psedudoscalar system. Then the transition amplitudes $t_{K^+\bar{K}^0\to K^+\bar{K}^0 }$ in particle basis can be related to the ones in isospin basis,
\begin{eqnarray}
t_{K^+\bar{K}^0\to K^+\bar{K}^0 }=t_{K\bar{K}\to K\bar{K} }.
\end{eqnarray}

\begin{figure}[htbp]
	\centering
	\includegraphics[scale=0.35]{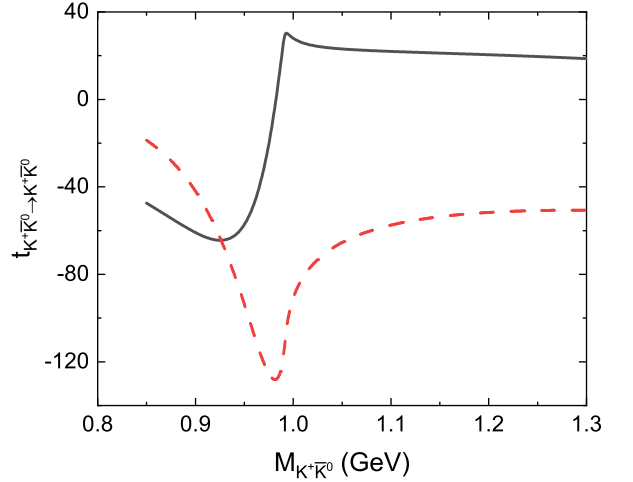}
\caption{Real (solid curve) and imaginary (red-dashed curve) parts of the transition amplitudes $t_{K^+\bar{K}^0\to K^+\bar{K}^0 }$ as a function of the invariant mass of the $K^+\bar{K}^0$ system.}	\label{fig:tkkkk}
\end{figure}

In Fig.~\ref{fig:tkkkk}, we show the real and imaginary parts of the transition amplitudes $t_{K^+\bar{K}^0\to K^+\bar{K}^0 }$ as a function of the invariant mass of the $K^+\bar{K}^0$ system. One can see that there is a clear structure around $980$ MeV, which is associated to the $a_0(980)$ resonance.

\subsection{The mechanism of $D^+_s \to \pi^0 (K^*\bar{K}^*)^+ \to K^0_SK^+ \pi^0 $ reaction}

\begin{figure}[tbhp]
    \centering
        \includegraphics[scale=0.49]{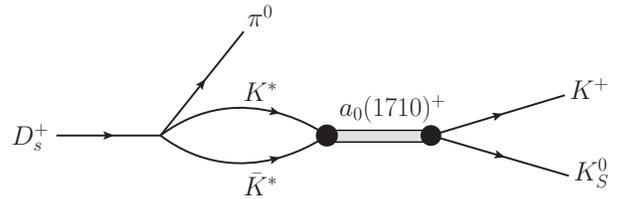}
    \caption{Feynman diagramn for the $D^+_s \to \pi^0 (K^*\bar{K}^*)^+ \to \pi^0 a_0(1710)^+ \to \pi^0 K^+K^0_S$ decay, where  $a_0(1710)$ is dynamically generated by the $K^*\bar{K}^*$ final-state interaction.}
  \label{fig:ma}
\end{figure}

Following Ref.~\cite{Zhu:2022wzk}, after the $\pi^0$ and $K^{*+} \bar{K}^{*0}$ pair produced, the final-state interactions of $K^{*+}$ and $\bar{K}^{*0}$ will lead to the dynamical generation of the $a_0(1710)^+$ state, which then decays into $K^+$ and $K^0_S$ in the final state. The rescattering diagram for the $D^+_s \to \pi^0 (K^*\bar{K}^*)^+ \to \pi^0 a_0(1710)^+ \to \pi^0 K^+K^0_S$ decay is shown in Fig.~\ref{fig:ma}. The decay amplitude can be written as
\begin{eqnarray}
&&\mathcal{M}_{a_0(1710)} = {\frac{V_{\pi^0K^{*+}\bar{K}^{*0}}}{2}}{\tilde{G}}_{K^{\ast}\bar{K}^{\ast}}(M_{K^+K^0_S}) \times \nonumber \\
&&\!\!\!\!\!\!\!{\frac{g_{K^\ast\bar{K}^\ast}g_{K\bar{K}}}{M_{K^+K_S^0}^2-M_{a_0(1710)^+}^2+iM_{a_0(1710)^+}\Gamma_{a_0(1710)^+}}}. \label{eq:ma}
\end{eqnarray}

It is worth mentioning that, in the present work, all the model parameters for the intermediate $a_0(1710)$  are determined in Ref.~\cite{Zhu:2022wzk}. Since  the parameters of both Set I and II can be used to  reproduce the experimental data on the $D^+_s \to \pi^+K^0_SK^0_S$ decay in  Ref.~\cite{Zhu:2022wzk}, we will use the values of Set I  in this work.

\subsection{Invariant mass distributions of the $D^+_s \to \pi^0K^+K_S^0$ decay}

With the above formalism and ingredients, we can write the total decay amplitude of $D^+_s \to \pi^0 K^+K^0_S$ as follows,
\begin{eqnarray}
 \mathcal{M} = \mathcal{M}_{K^{\ast+}}+\mathcal{M}_{\bar{K}^{\ast0}}+\mathcal{M}_{a_0(980)}+\mathcal{M}_{a_0(1710)}.
 \label{eq:m}
\end{eqnarray}
The amplitude of Eq.~\eqref{eq:m} depends on two invariant masses $M_{K^+K_S^0}$ and $M_{{\pi}^0K^+}$.

Then, the double differential decay width for  $D^+_s \to \pi^0 K^+K^0_S$ is given by~\footnote{We take $|K^0\rangle = \frac{1}{\sqrt{2}}
	(|K^0_S\rangle + |K^0_L\rangle)$ and $|\bar{K}^0\rangle =
	\frac{1}{\sqrt{2}} (|K^0_S\rangle - |K^0_L\rangle)$, where we have
	neglected the effect of $CP$ violation.}
\begin{eqnarray}
     &&\frac{d^2\Gamma}{dM_{K^+K_S^0}{dM_{{\pi}^0K^+}}}=\frac{M_{K^+K_S^0}M_{{\pi}^0K^+}}{128\pi^3m_{D_s^+}^3}  \left(|\mathcal{M}_{K^{\ast+}}|^2\right. \nonumber \\
     &&   +\left.|\mathcal{M}_{\bar{K}^{\ast0}}|^2 + |\mathcal{M}_{a_0(980)}|^2 +|\mathcal{M}_{a_0(1710)}|^2\right), \label{eq:dgammadm12dm23}
\end{eqnarray}
where their interference terms are neglected, since all these coupling constants are assumed to be real and positive, as discussed above.

Finally, one can easily obtain the invariant mass distributions ${d\Gamma}/{dM_{K^+K_S^0}}$ and ${d\Gamma}/{dM_{{\pi}^0K^+}}$, by integrating Eq.~\eqref{eq:dgammadm12dm23} over each of the invariant mass variables. For example, for a given value of $M_{K^+K_S^0}$, the upper and lower limits for $M_{\pi^0K^+}$ are fixed as~\cite{Workman:2022ynf},
\begin{eqnarray}
\left(M_{\pi^0K^+}^2\right)_\text{max} &= &\left(E_{\pi^0}^\ast+E_{K^+}^\ast\right)^2 -  \nonumber \\
    && \left(\sqrt{E_{\pi^0}^{\ast2}-m_{\pi^0}^2}-\sqrt{E_{K^+}^{\ast2}-m_{K^+}^2}\right)^2 \nonumber \\
\left(M_{\pi^0K^+}^2\right)_\text{min} &=&\left(E_{\pi^0}^\ast+E_{K^+}^\ast\right)^2 -  \nonumber \\
    &&\left(\sqrt{E_{\pi^0}^{\ast2}-m_{\pi^0}^2}+\sqrt{E_{K^+}^{\ast2}-m_{K^+}^2}\right)^2, \nonumber
\end{eqnarray}
here $E_{\pi^0}^\ast$ and $E_{K^+}^{\ast}$ are the energies of $\pi^0$ and $K^+$ in the $K^+K_S^0$ rest frame, respectively,
\begin{align}
    &E_{\pi^0}^\ast=\frac{m_{D_s^+}^2-M_{K^+K_S^0}^2-m_{\pi^0}^2}{2M_{K^+K_S^0}},  \nonumber \\
    &E_{K^+}^\ast=\frac{M_{K^+K_S^0}^2-m_{K_S^0}^2+m_{K^+}^2}{2M_{K^+K_S^0}}.
\end{align}

\section{Results and Discussion} \label{sec:Results}

With these above ingredients, we calculate the invariant mass distributions of the $D^+_s \to \pi^0 K^+ K^0_S$ decay. In the numerical analysis, we have three free parameters:~(1) the factor $V_{\pi^0K^{*+}\bar{K}^{*0}}$ of Eq.~\eqref{eq:ma} for the weak and hadronization strength related to the production of intermediate $\pi^0 (K^*\bar{K}^*)^+$ at tree level; (2) $V_{\pi^0K^+\bar{K}^0}$ for the weight of the contribution from the intermediate $a_0{(980)}^+$ state; (3) a global normalization factor $C$, which is needed to normalize the theoretical invariant mass distributions to the events obtained by the BESIII Collaboration. 

For the parameter $V_{\pi^0K^{*+}\bar{K}^{*0}}$ ($V_P$ in Ref.~\cite{Zhu:2022wzk}), we have two choices. One (Set I) is just taking the value from the previous work in Ref.~\cite{Zhu:2022wzk}, which is determined from the contribution of $a_0(1710)$ to the $D^+_s \to \pi^+K^0_SK^0_S$ reaction. The other one (Set II) is determined from the new experimental data on the $D^+_s \to \pi^0K^+K_S^0$ reaction. While for $V_{\pi^0K^+\bar{K}^0}$ and $C$, they will be determined from the current $D^+_s \to \pi^0K^+K_S^0$ data. These above model parameters are shown in Table \ref{table:parameters}.

\begin{table}[htbp]
\renewcommand\arraystretch{1.2}
\caption{Model parameters.}	\label{table:parameters}
	\begin{tabular}{ccc}
		\hline\hline  
	~~~~~~~Parameters~~~~~~~ & ~~~~~~~Set I~~~~~~~  & ~~~~~~~Set II~~~~~~~  \\ \hline
		$C$  & $3.62\times10^{14}$    & $3.21\times 10^{14}$    \\
		$V_{\pi^0K^{*+}\bar{K}^{*0}}$ & $1.69\times10^{-4}$  (Ref.~\cite{Zhu:2022wzk})   & $2.4\times10^{-4}$  \\
		$V_{\pi^0K^+\bar{K}^0}$  & $7.5\times10^{-5}$    & $8.0\times10^{-5}$   \\
		\hline\hline
	\end{tabular}
\end{table}

\begin{figure}[tbhp]
\begin{center}
\includegraphics[scale=0.32]{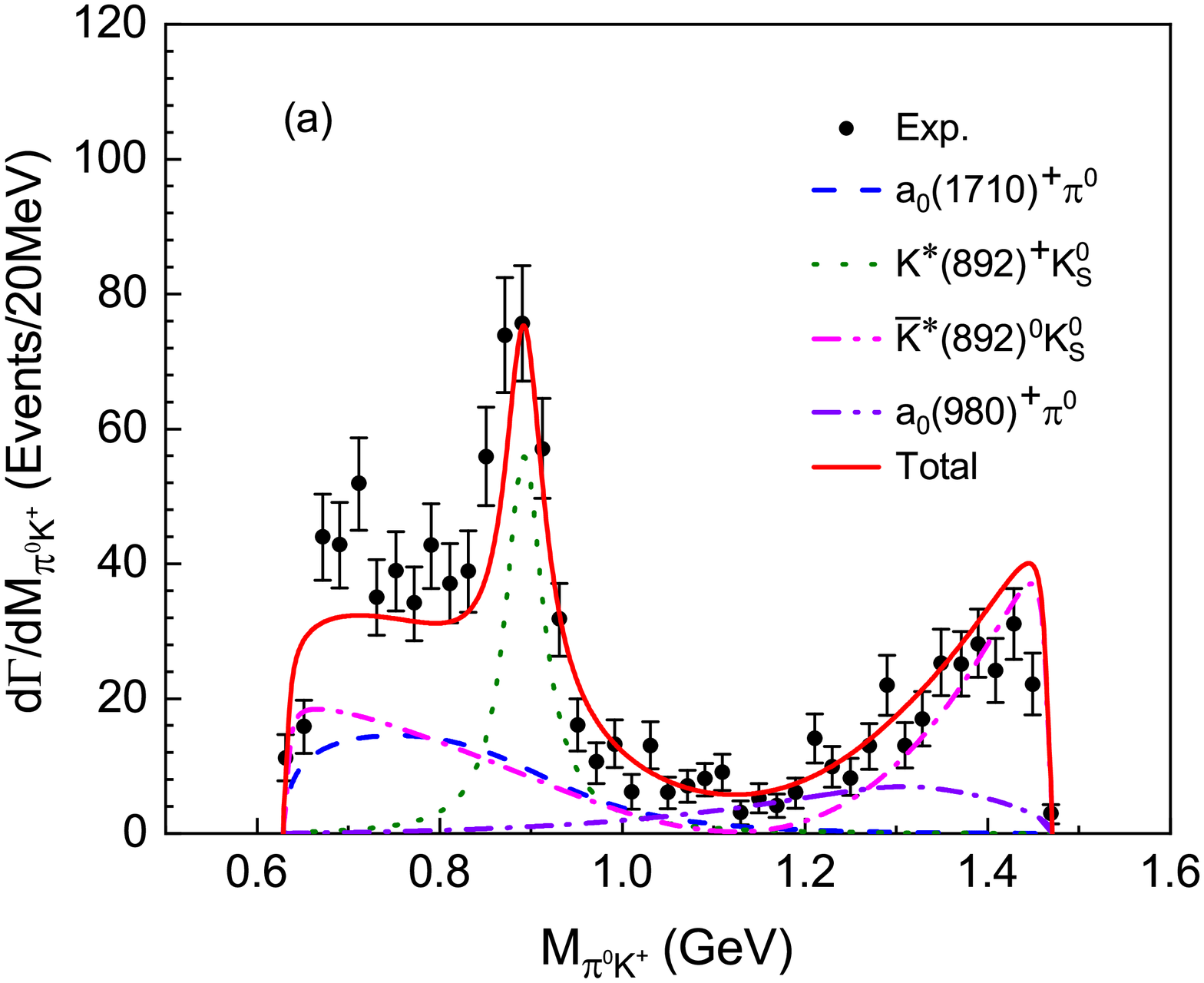}
\includegraphics[scale=0.32]{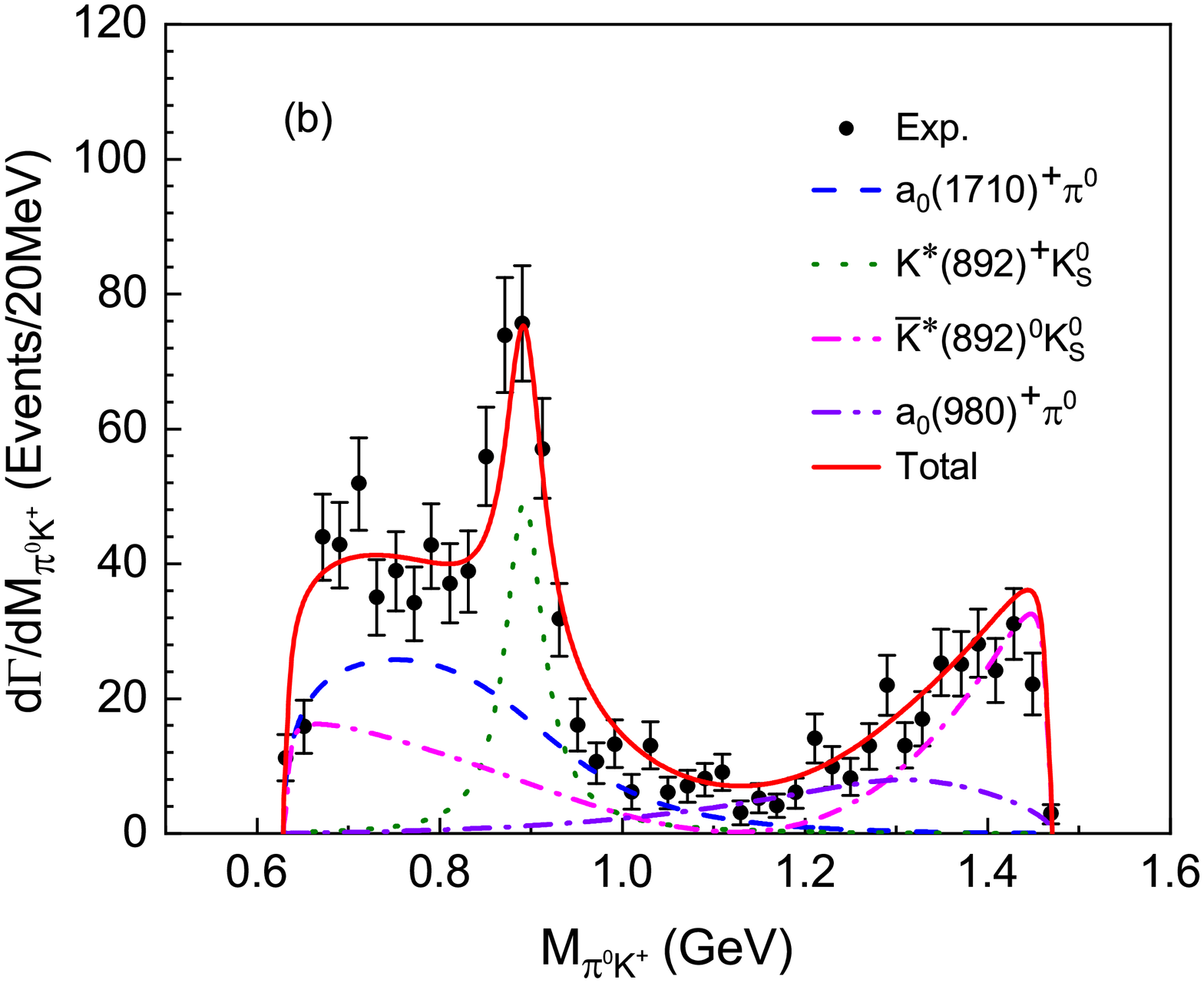}
\end{center}
\vspace{-0.5cm}
\caption{Invariant mass distributions of $\pi^0K^+$ for the $D_s^+\rightarrow\pi^0K^+K_S^0$ decay, compared with the experimental data taken from Fig.~2(c) of Ref.~\cite{BESIII:2022wkv}. (a)  the results of Set I, and (b)  the results of Set II.} \label{fig:pi0kz}
\end{figure}

We firstly show the theoretical results in Fig.~\ref{fig:pi0kz} for the $\pi^0K^+$ invariant mass distributions, where the red-solid curves stand for the total contributions. While the blue-dashed, green-dotted, pink-dot dashed, and purple-dot dashed curves correspond to the contributions from the intermediate states $a_0(1710)^+$, $K^{\ast+}$, $\bar{K}^{\ast 0}$, and $a_0(980)^+$, respectively. The total results have been normalized in both cases to the peak of the $\pi^0 K^+$ invariant mass distributions. The corresponding values of $C$ are $3.62 \times 10^{14}$ and $3.21\times 10^{14}$ for Set I and Set II, respectively. It is shown that our results can describe well the experimental data. Once again, as for the $D^+_s \to \pi^+K^0_SK^0_S$ reaction~\cite{Zhu:2022wzk}, one can see that the contribution of the $a_0(1710)$ state is crucial to a satisfactory description of the $\pi^0K^+$ invariant mass distributions. Furthermore, the higher tail of the $\pi^0 K^+$ line shape is mostly from the reflection of the $a_0(980)$ resonance, which implies that the contribution from the vector meson $K^{*+}(1410)$ should be small and negligible.

\begin{figure}[tbhp]
\begin{center}
\includegraphics[scale=0.32]{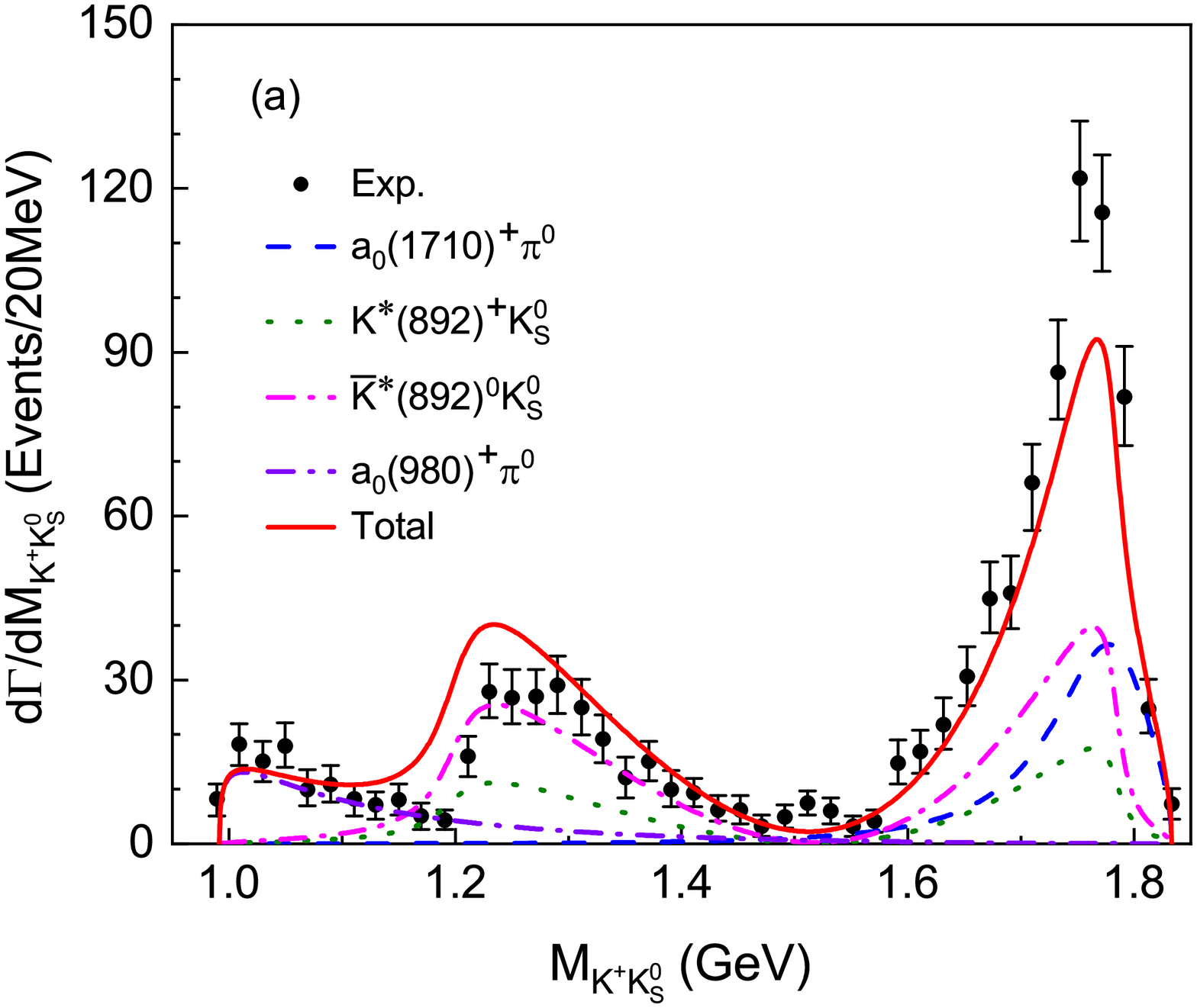}
\includegraphics[scale=0.32]{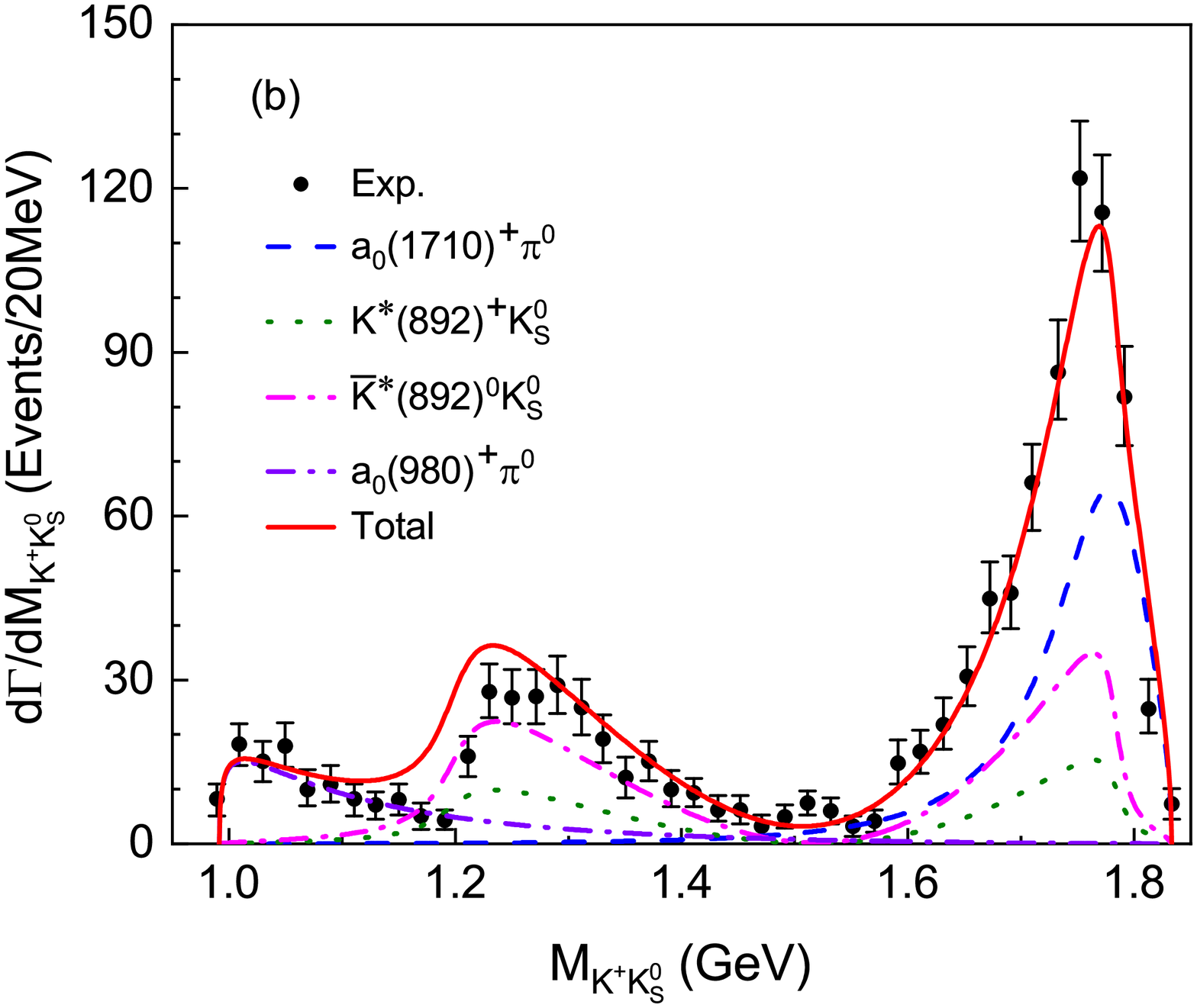}
\end{center}
\vspace{-0.5cm}
\caption{Invariant mass distributions of $K^+ K_S^0$ for the $D_s^+\rightarrow\pi^0K^+K_S^0$ decay, compared with the experimental data taken from Fig.~2(a) of Ref.~\cite{BESIII:2022wkv}. (a)  the results of Set I, and (b)  the results of Set II.}\label{fig:ksokz}
\end{figure}

Next, in Fig.~\ref{fig:ksokz}, we show the theoretical results on the $K^+K^0_S$ invariant mass distributions. From Fig.~\ref{fig:ksokz} (a), one can see that with the value $1.69\times10^{-4}$ of $V_{\pi^0K^{*+}\bar{K}^{*0}}$ determined in Ref.~\cite{Zhu:2022wzk}, the peak of $a_0(1710)^+$ can not be well reproduced, and the theoretical results are lower than the experimental data. However, by adjusting the value of $V_{\pi^0K^{*+} \bar{K}^{*0}}$, we can reproduce the $a_0(1710)^+$ peak very well, which is shown in Fig.~\ref{fig:ksokz} (b). In both cases, the bump for the $a_0(980)^+$ state can be well explained, and the corresponding values for $V_{\pi^0 K^+\bar{K}^0}$ are $7.5\times10^{-5}$ and $8.0\times10^{-5}$, respectively. These two values are very similar to each other. In addition, the $K^{*+}$ and $\bar{K}^{*0}$ mesons contribute significantly to the peak reagion of the $a_0(1710)^+$ resonance.

\begin{figure}[tbhp]
\begin{center}
\includegraphics[scale=0.32]{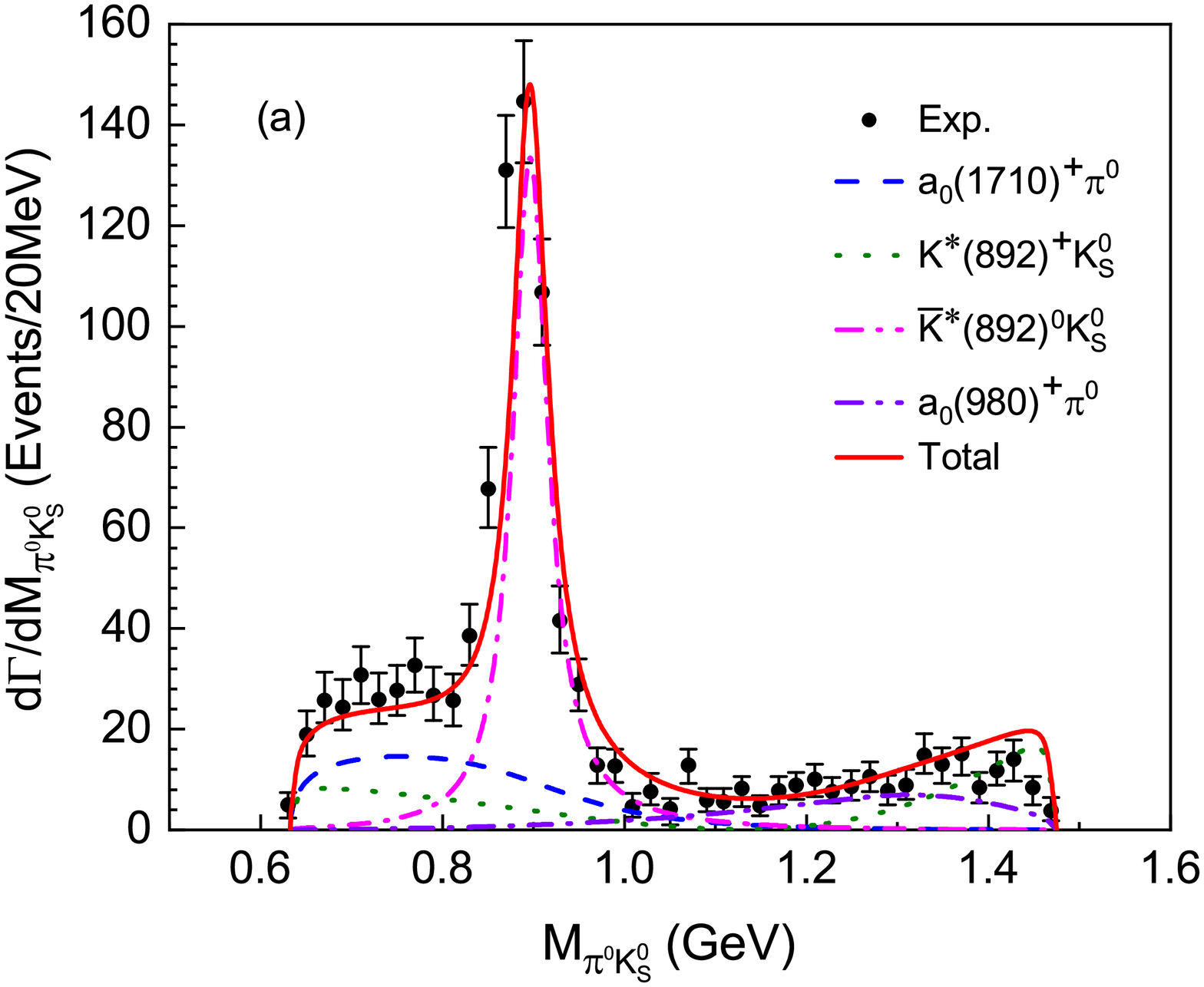}
\includegraphics[scale=0.32]{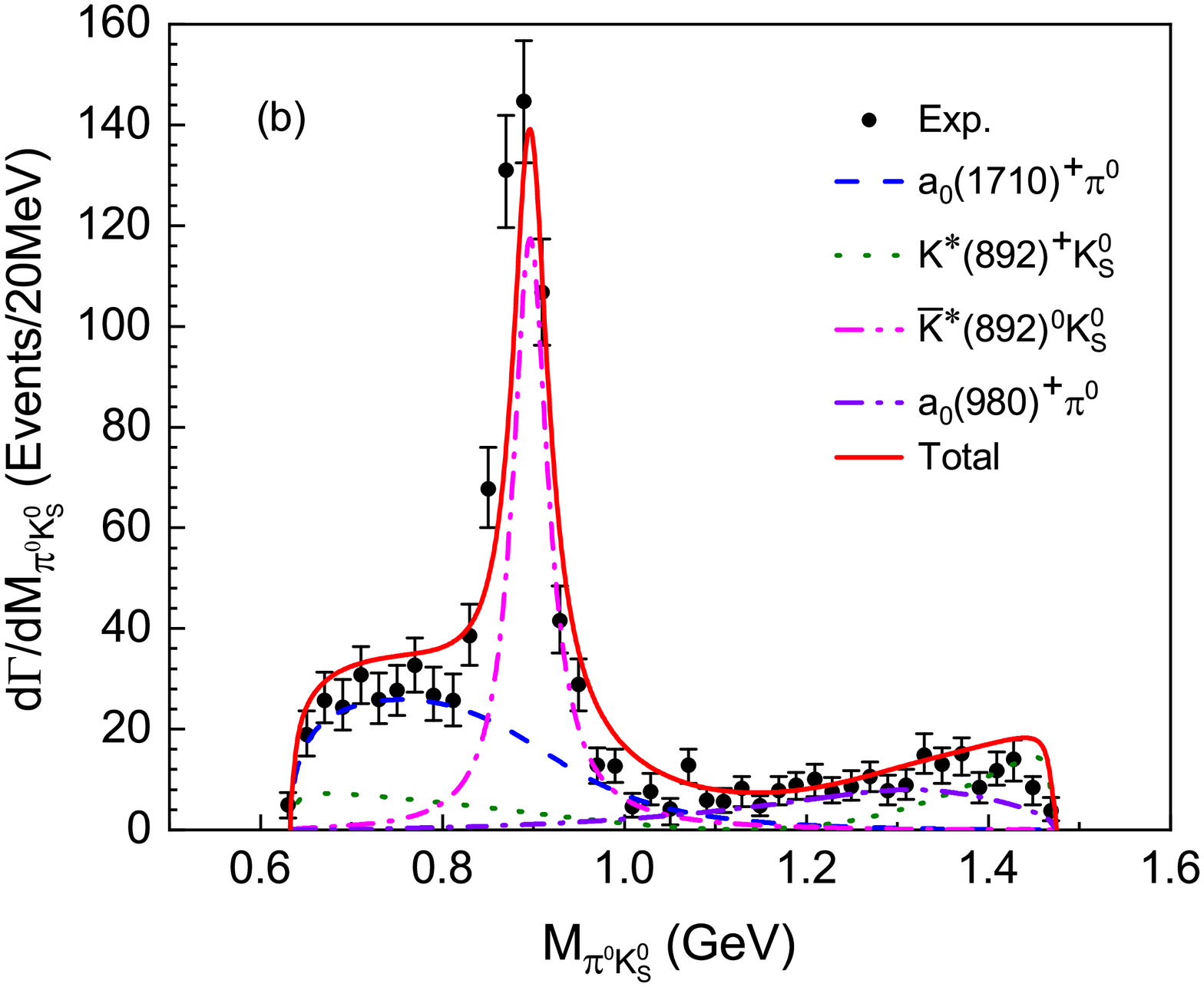}

\end{center}
\vspace{-0.5cm}
\caption{Invariant mass distributions of $\pi^0K_S^0$ for the $D_s^+\rightarrow\pi^0K^+K_S^0$ decay, compared with the experimental data taken from Fig.~2(b) of Ref.~\cite{BESIII:2022wkv}. (a)  the results of Set I, and (b)  the results of Set II.} \label{fig:pi0ks0}
\end{figure}

Finally, in Fig.~\ref{fig:pi0ks0}, the theoretical results for the $\pi^0 K^0_S$ invariant mass distributions are shown. It is easily seen that the experimental data can be well reproduced in both cases. Again, the higher tail of the $\pi^0K^0_S$ line shape is from the $a_0(980)$ resonance, rather than the vector meson $\bar{K}^{*0}(1410)$, which is needed in the analysis by the BESIII Collaboration~\cite{BESIII:2022wkv}.

It is worth mentioning that, in the present work, we do not consider the interference effects among the $\mathcal{M}_{K^{\ast+}}$, $\mathcal{M}_{\bar{K}^{\ast0}}$, $\mathcal{M}_{a_0(980)}$, and $\mathcal{M}_{a_0(1710)}$ amplitudes. We note that in Ref.~\cite{Ling:2021qzl} the $a_0(980)$ state was studied in the $D^+_s \to \pi^0K^+\bar{K}^0$ decay by including the contributions from the triangle diagrams involving the intermediate $K^*$ and $\rho$ resonances, and then the $K\bar{K}$ and $\pi\eta$ pairs fuse to generate the $a_0(980)$ state. It is found that the signal of the $a_0(980)$ state is also clearly  seen~\cite{Ling:2021qzl}. In this work, for the production of $a_0(980)$, we rely on the production of the $\pi^0K^+\bar{K}^0$, where all particles are produced in $S$-wave. Yet, it is easy to see that the
mechanism proposed in Ref.~\cite{Ling:2021qzl} should be suppressed due to the highly off-shell
effect of the $K^*$ propagator when the $K^+\bar{K}^0$ invariant mass is close to the $a_0(980)^+$ mass. Furthermore, to include such contributions  more free parameters are needed. In addition, we have checked the $S$-wave final-state interactions of $\pi^0$ and $K^+$ ($\bar{K}^0$), and found that these two contributions are much smaller compared with the contributions of the $K^{*+}$ and $\bar{K}^{*0}$. Nevertheless, the model proposed in the present work can give a reasonable description of the experimental data for the $D^+_s \to \pi^0 K^+ K^0_S$ decay and constitutes a further theoretical effort in studying the roles of the $a_0(980)$ and $a_0(1710)$ states in the relevant reaction.

\section{Summary} \label{sec:Conclusions}

In the present work, we investigated the Cabibbo-favored process $D_s^+\rightarrow\pi^0K^+K_S^0$, taking into account the $S$-wave $(K^{\ast}\bar{K}^{\ast})^+$ final-state interaction within the chiral unitary approach, where the $a_0(1710)^+$ state is dynamically generated. In addition, the contributions from the tree diagrams of $K^{\ast+} \to \pi^0K^+$ and $\bar{K}^{\ast0}\to\pi^0K^0_S$ were also considered. While for the $K\bar{K}$ final-state interaction, we took into account the $a_0(980)^+$ state, which is a dynamically generated resonance from the coupled-channel pseudoscalar-pseudoscalar interaction. Considering all these contributions, we calculated the three invariant mass distributions of $K^+K^0_S$, $\pi^0K^+$, and $\pi^0K_S^0$, to which  $a_0(1710)^+$, $a_0(980)^+$, $K^{\ast+}$ and $\bar{K}^{\ast0}$ contribute, respectively. 

We showed that the new experimental data on the invariant mass distributions of  ${d\Gamma}/{dM_{{\pi}^0K^+}}$, ${d\Gamma}/{dM_{K^+K_S^0}}$, and ${d\Gamma}/{dM_{{\pi}^0K^0_S}}$ measured by the BESIII Collaboration~\cite{BESIII:2022wkv} can be well reproduced. We found that the vector meson $K^*(892)$ plays a crucial role in the $a_0(1710)^+$ peak region. On the other hand, the contributions from $K^*(1410)$ were not needed. The higher tail of the $\pi K$ line shapes can be well reproduced due to the reflection effect of the $a_0(980)$ resonance.

We would like to stress that the decay of $D_s^+ \rightarrow \pi^0 K^+ K_S^0$ is a good platform to study the isospin one scalar mesons, since the final $K^+K_S^0$ system is in pure isospin $I=1$. The experimental measurements on the $D^+_s \to \pi^0K^+K^0_S$ reaction by the BESIII Collaboration support the $K^*\bar{K}^*$ nature of the $a_0(1710)$ state, and the $a_0(980)$ state is a dynamically generated state from the coupled-channel interactions of psedudo-scalar meson and pseudo-scalar meson. Furthermore, thanks to the important role played by the scalar mesons $a_0(980)$ and $a_0(1710)$  in the
$D^+_s \to \pi^0K^+K^0_S$ reaction, the accurate data on this reaction
can be used to improve our knowledge about these scalar mesons.

\section*{Acknowledgments}

This work is partly supported by the National Natural Science Foundation of China under Grants No. 12075288, No. 11735003, No. 11975041, No. 11961141004, No. 11961141012, and No. 12192263. This work is supported by the Natural Science Foundation of Henan under Grant No. 222300420554,  the Project of Youth Backbone Teachers of Colleges and Universities of Henan Province (2020GGJS017), the Youth Talent Support Project of Henan (No. 2021HYTP002), and the Open Project of
Guangxi Key Laboratory of Nuclear Physics and Nuclear Technology, No. NLK2021-08.


\end{document}